\documentclass[conference,compsoc]{IEEEtran}
\usepackage{graphicx}
\usepackage[caption=false,font=footnotesize,labelfont=sf,textfont=sf]{subfig}
\usepackage{url}
\usepackage{color}
\usepackage{colortbl}
\usepackage{amssymb,array}

\ifCLASSOPTIONcompsoc
 \usepackage[nocompress]{cite}
\else
 \usepackage{cite}
\fi
\ifCLASSINFOpdf
\else
\fi

\begin{document}
%
\title{Development of a cyber risk assessment tool for Irish small business owners }

\author{\IEEEauthorblockN{Miriam Curtin\IEEEauthorrefmark{1},
Brian Sheehan\IEEEauthorrefmark{2},
Melanie Gruben\IEEEauthorrefmark{3}, 
Nikoletta Kozma\IEEEauthorrefmark{4},
Gillian O'Carroll\IEEEauthorrefmark{5} and
Hazel Murray\IEEEauthorrefmark{6}}
\IEEEauthorblockA{Munster Technological University\\Ireland\\
}}


%


\maketitle

\begin{abstract}
Small and medium enterprises (SMEs) are increasingly vulnerable to cyber threats due to limited resources and cybersecurity expertise, in addition to an increasingly hostile cyber threat environment at national and international levels. This study aims to improve the cyber resilience amongst SMEs by developing a national risk assessment tool. This research is guided by three key questions: 1. What current international SME risk assessment tools are available and supported or endorsed by national cybersecurity centres? 2. How can a risk assessment tool be created that is accessible to SME owners with little to no cybersecurity knowledge? 3. What are the key areas of cybersecurity risks for SMEs? To answer these questions, a comprehensive review of existing risk assessment tools was carried out. Through iterative collaboration with SMEs, the development of a user-friendly tool that simplifies risk for non-expert users was made possible. 
\end{abstract}


\section{Introduction}
Small and medium enterprises (SMEs) are among the most vulnerable to cybersecurity risks, characterized by a lack of cyber resilience~\cite{benz2020calculated} and low expenditure on cybersecurity~\cite{ng2022sps2022}. Most SMEs also lack specialists in data security to deal with security issues, perform regular backups, and keep their programs and software updated, this gap often results in their vulnerability to exploitation~\cite{mitrofan2020determining}. 

Despite playing a crucial role in the economy, SMEs face significant challenges due to limited cybersecurity knowledge and resource constraints, making them particularly susceptible to cyberattacks~\cite{koeze2017designing}. SMEs also lack the capacity to abide by regulatory standards, such as, the general Data Protection Regulations (GDPRs), that have been designed to enhance organizational cybersecurity practices and cultures~\cite{chaudhary2023quest}. A risk assessment tool can assist SMEs in identifying and mitigating areas of cyber risk that they encounter.

In the paper titled ``Designing a cyber risk assessment tool for small to medium enterprises''~\cite{koeze2017designing}, the review of existing literature concluded that there are few to no effective cybersecurity solutions available for SMEs. This author’s proposed solution involved adapting the TREsPASS model~\cite{Trespass}, slimming it down to fit the needs of an SME. However, after conducting interviews with an SME and cybersecurity expert, it became evident that the requirements for their tool needed further expansion. This finding underscored the importance of collaborating with SMEs during the development of our risk assessment tool.

Additionally, to ensure the tools are useful and tailored to the SMEs' unique needs it is crucial to involve SMEs in the designing of the cybersecurity tools. Implementing this method also encourages SMEs to take ownership, making them more likely to implement and consistently use these tools~\cite{cyberreadiness}. Our study aims to involve SMEs in many different ways, from stakeholder engagement during the ideation stage to piloting via think aloud exercises and finally roll our and feedback in focus groups. This provided them an opportunity to test the tool and raise any questions or concerns they have, we received valuable feedback from these activities.

The core goal of this study is to enhance the cyber resilience of SMEs by helping them to identify their specific cybersecurity needs. This will be achieved through the development of a national cyber risk assessment tool for SMEs. To this end, we pose three research questions: 

\textbf{RQ1: What current international SME risk assessment tools are available and supported or endorsed by national cybersecurity centres? }

Analysing the existing cyber risk assessment tools is vital.  National cybersecurity centres often set the standard for best practices in cyber risk management and their endorsed tools will reflect this.  By analysing their endorsed risk assessment tools, we can identify strengths and limitations, and use these to develop our own national risk assessment tool specifically tailored to Irish small business owners.  

\textbf{RQ2: How can a risk assessment tool be created that is accessible to SME owners with little to no cybersecurity knowledge? }

As SMEs tend to not have expertise in cybersecurity, it is essential that the tool uses non-technical language.  SMEs are typically constrained by time and budget resources, making it crucial that the tool considers these factors.  To cater for these factors, the risk assessment tool must be easy to understand and use, should not take a significant amount of time to complete, and should be free of charge.

\textbf{RQ3: What are the key areas of cybersecurity risks for SMEs? }

Identifying the key areas of cybersecurity risks for SMEs will guide future research and supports for small businesses. By analysing the results of the risk assessment the researchers present the top cybersecurity risk areas across SMEs.

By addressing these research questions, this study aims to contribute to the cyber resilience of SMEs.  The outcomes include providing SMEs with a practical and user-friendly risk assessment tool, enabling them to better safeguard their businesses and reduce their cybersecurity risks.

\section{Literature review}
SMEs make up 99.8\% of all Irish businesses and 90\% of businesses worldwide and account~\cite{statista-SME,worldbank}. They account for 70\% of Irish employment and 50\% of employment worldwide. However, SMEs often have the attitude that they wont be targeted by cyber attacks~\cite{wilson2023won}. SMEs are also more likely to be in the early stages of development, and less likely to be fully established in all aspects of operations~\cite{scott1987five}. SMEs lack the resources that larger organisations have to protect against cyberattacks~\cite{sandalgaard2018budget}. Historically, SMEs have faced barriers around upgrading technology when compared to larger businesses, due to SMEs having less scope for cutting-edge cybersecurity~\cite{a2014mobile}. Cyber security advice is often tailored to large businesses, assuming consistent documentation processes~\cite{heidt2019investigating} and a skilled workforce~\cite{onwubiko2007managing}. Lack of employee training is one of the biggest cybersecurity risks to SMEs~\cite{sabillon2021effective}. Larger organisations are more likely to have designated and educated cyber security staff~\cite{zec2015cyber} and overall cyber resilience skillsets may be lacking~\cite{mijnhardt2016organizational}. Leaders in SMEs may be more involved in operations and day-to-day duties, rather than focused on cyber security best practices~\cite{bhattacharya2013evolution}. SMEs are also more likely to lack knowledge on securing organisational data~\cite{gafni2019invisible}. Compounding all this, cyberattacks on SMEs can cause catastrophic financial impact~\cite{chen2022information, chidukwani2022survey}.

Rather than simply a traditional cybersecurity approach, some evidence promotes the benefits of cyber resilience - a holistic approach to cyber security which includes multiple networks, safe-to-fail versus fail-safe approaches, and the importance of not interrupting business~\cite{linkov2013resilience,carias2021cyber}. Research has shown that small businesses are more likely to take action to improve cybersecurity when technical tools are simple, cybersecurity expertise is easy to access, and educational materials are easily understood~\cite{eilts2020empirical}. One benefit that SMEs have is greater agility, allowing them to pivot business practices or organisational response more quickly to adjust their cyber security approach~\cite{caldwell2015securing}.

Many leaders in SMEs believe cyber security software is sufficient to protect against threats~\cite{brar2018cybercrimes}, which is not necessarily supported by cyber security researchers~\cite{kellaris2016generic}. Many SMEs do not view upgrading cyber security as
crucial~\cite{kljuvcnikov2019information} and past research has found that cyber security culture is less present within SMEs~\cite{santos2016importance} and that SMEs often have weak understanding of information systems~\cite{gafni2023experts}. SMEs are more likely to mix personal and professional devices for use in the business, making a less streamlined IT system for implementing standardized cyber security protocol~\cite{tam2021good}. Similarly, past reports have found that the design of small business IT architecture can negatively impact their cyber security more than it would for a large businesses~\cite{osborn2017small}. SMEs have historically been more likely to store information in a hybrid (partly analog) fashion, but information security services do not necessarily accommodate this~\cite{Dhoha18}. Also most companies do not understand the role of insurance coverage in the case of a cyberattack~\cite{satariano2019big}. 

At a glance, cyber security within a business is not necessarily easily measured, and SMEs must account for return on investment~\cite{goode2018expert,paul2019socially}. Research has also found that decision makers in small businesses may find cyber security advice overwhelming~\cite{bell2017cybersecurity} and that SME decision-makers may not know how to assess their business for cyber security~\cite{udofot2020factors}. Indeed, while there are many cyber security assessment frameworks already~\cite{cybersecurity2018framework,linkov2013resilience,mohammed2019cybersecurity}, these frameworks are not tailored for SMEs as they often have hundreds of specific policies which are not always relevant to SMEs. Customised analysis tools for cyber vulnerability may therefore benefit SMEs~\cite{bernik2016measuring}. SMEs are more likely to lack internal knowledge of cyber risk assessments~\cite{bada2019developing}, but past research has shown the benefits of SME-oriented risk assessments~\cite{carlton2019mitigating,carias2021cyber,al2023risk}.

Previous researchers who created a risk assessment tool used frameworks such as NIST or ISO to guide them~\cite{koeze2017designing}. While some studies exclusively used one framework~\cite{armenia2021dynamic} others combined frameworks~\cite{carias2021cyber} or developed entirely new ones~\cite{tam2019macra}. Using a framework can ensure that the risk assessment tool covers key areas of risk for an SMEs. Our risk assessment is informed by the topics covered in international risk assessment tools, academic literature, the NIST Cybersecurity Framework v2.0~\cite{NISTCSF} and insights from SMEs.

The aim of this research is to develop a practical and valuable risk assessment tool tailored to small businesses. Therefore the emphasis is on collaborative iterative design with small business owners and grounded research in existing real-world solutions.




\section{Methodology}\label{sec:meth}
This research involved a three-part study to develop a risk assessment tool for Irish small businesses. The study began with an analysis of existing international risk assessment tools. Based on this analysis, we developed our own risk assessment tool, which was then piloted with SME owners or managers using a think aloud activity. Updates were made to the original risk assessment tool based on SME owners' feedback. The next version of the risk assessment tool was then used in two different focus groups. Further changes were made in accordance with the focus groups feedback leading to the creation of the final version of the risk assessment tool. 

\subsection{International Risk Assessment Tools}
The following inclusion criteria were applied for our survey of existing risk assessment tools: 

Requirements: The risk assessment tool must be supported or endorsed by the country's National Cybersecurity Centre. 

Country Selection: We aimed to analyze a risk assessment tool from at least one country from each continent. However, a risk assessment tool endorsed by the countries' own national cybersecurity centres was not easily found for Africa or South America. The six countries that were included in the final selection are: Belgium \cite{belgianRAT}, Singapore \cite{SingaporeRAT}, Australia \cite{AustralianRAT}, Finland \cite{FinlandRAT}, UK \cite{UKRAT}, and US\cite{USRAT}. 

Analysis: The risk assessment tools were analysed under the following headings: time taken to complete, number of questions, number of questions containing jargon, ease of answering questions (based on this scale: 1=need advanced cybersecurity knowledge, 2=IT personnel, 3=regular owner, 4=regular employee), topics covered, general comments, and when the webpage was last updated. These were the core factors considered when designing our risk assessment tool. 

\subsection{Think-Aloud Interviews }
21 SME owners or managers who participated in stakeholder interviews were contacted again with an invitation to participate in this study. 14 of these participants accepted the invitation. 

Instructions: The participants were briefed about the project, asked to read the information sheet and then fill in a consent form. Participants were then asked to fill in the risk assessment tool, and the feedback questions at the end, while voicing their thoughts and opinions throughout. 

Methodology: The interviews were completed one-on-one on Microsoft Teams with a researcher. Notes were taken by the researcher as the participant voiced their thoughts. which were then uploaded to a word document. 

Feedback analysis: One researcher was assigned to making the changes to the risk assessment tool. If a change was made, the note was highlighted yellow. If the change was not made, the note was highlighted green along with an explanation as to why the feedback was not taken on. The risk assessment tool was updated after every three interviews to allow for the incorporation of feedback in manageable increments. This iterative approach enabled the researchers to make continuous improvements based on participant input, ensuring that the tool evolved to better meet the needs of SME owners. Additionally, this method allowed for the identification of recurring issues or suggestions, which could be addressed promptly, leading to more refined and effective subsequent versions. This iterative process also facilitated the collection of feedback on the revised versions, providing further insights and validation of the changes made. 

\begin{figure*}[!t]
\centering
\subfloat[Think Aloud participants represented 8 different business sectors]{\includegraphics[width=2.5in]{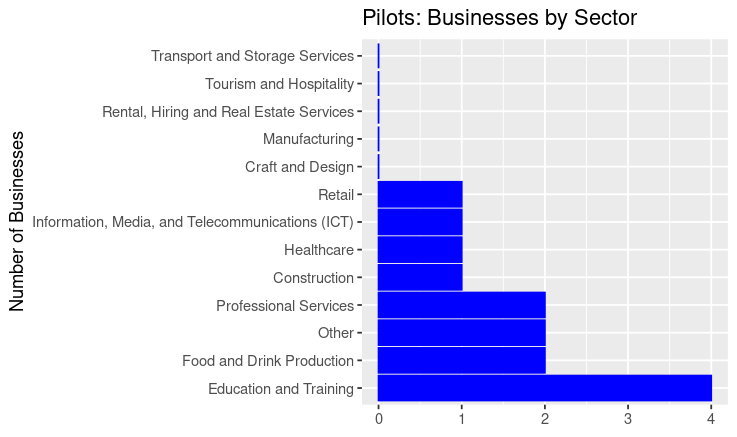}%
\label{pilot-sector}}
\hfil
\subfloat[Think aloud participants were owners of micro (0-9 employees), small (10-49 employees) and medium (50-249 employees) businesses.]{\includegraphics[width=2.5in]{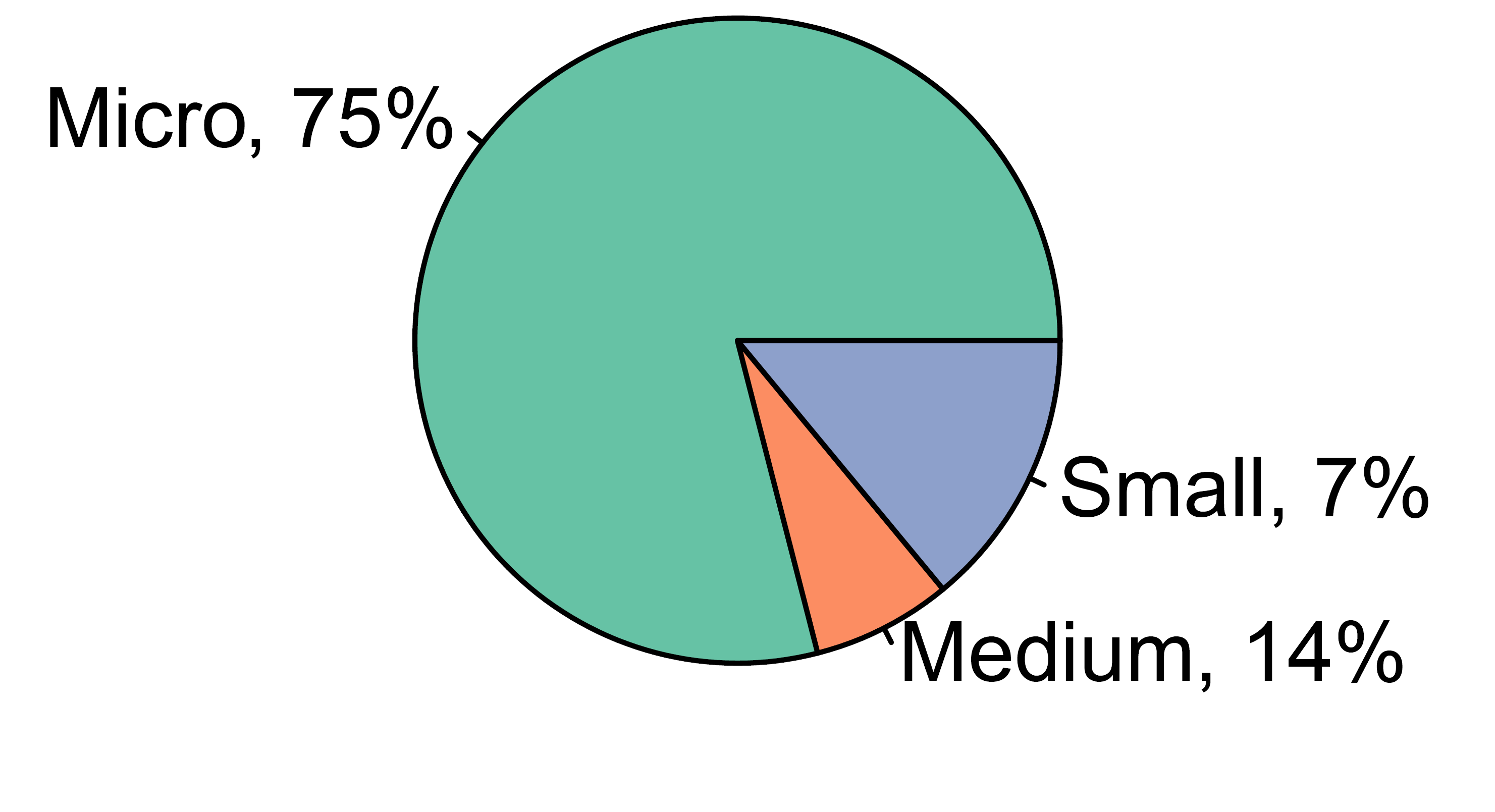}%
\label{pilot-size}}
\caption{Demographics of the businesses represented in the Think Aloud study.}
\label{think-aloud-demog}
\end{figure*}

Demographics: A total of 8 females and 6 males took part in this exercise. They represented 8 different sectors, with company sizes ranging from 1 to 185 employees. This data is represented in Figure~\ref{pilot-sector} and Figure~\ref{pilot-size}. 

\subsection{Focus Groups}
The research team worked with two focus groups that were conducted. The first focus group took place in Culture City, a co-working office space in Cork City, with 6 participants. The second focus group was organized in collaboration with the Kerry Local Enterprise Office, with 14 participants. The participants for each focus group were recruited with the help of the hosts and through advertising the opportunity to use the free cyber risk assessment tool. 

Instructions: The participants were briefed about the project, asked to read the information sheet and then fill in a consent form. They were then asked to fill in the risk assessment tool and the feedback questions at the end. Once everyone had completed the risk assessment tool, a discussion with the focus group was conducted. 

Methodology: A total of 15 participants completed the risk assessment tool in two different focus groups. They were given 10 minutes to complete the risk assessment tool, a guideline time from the think-aloud interviews. Feedback from the participants was collected both in written form (in the feedback section of the risk assessment tool) and orally during the group discussion after everyone completed the risk assessment tool. 

Feedback Analysis: The focus groups both had discussions after completing the risk assessment tool. No suggestions came from the discussions, just feedback on how they found the risk assessment tool. The written feedback provided some suggestions. 

\begin{figure*}[!t]
\centering
\subfloat[Focus group participants represented 9 different business sectors]{\includegraphics[width=2.5in]{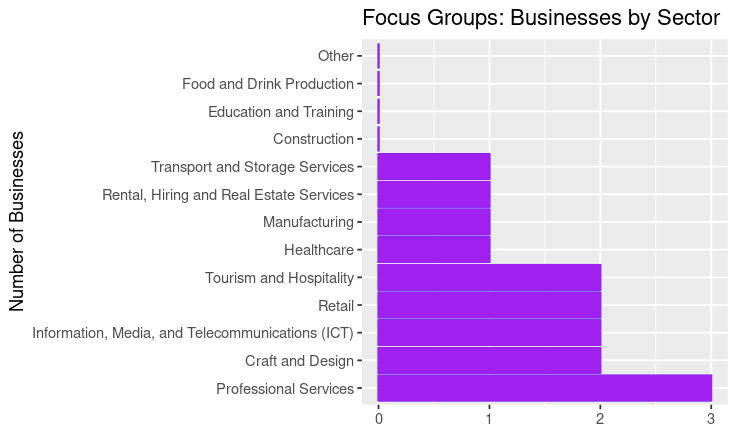}%
\label{FG-sector}}
\hfil
\subfloat[Focus group participants were owners of micro (0-9 employees), small (10-49 employees) and medium (50-249 employees) businesses.]{\includegraphics[width=2.5in]{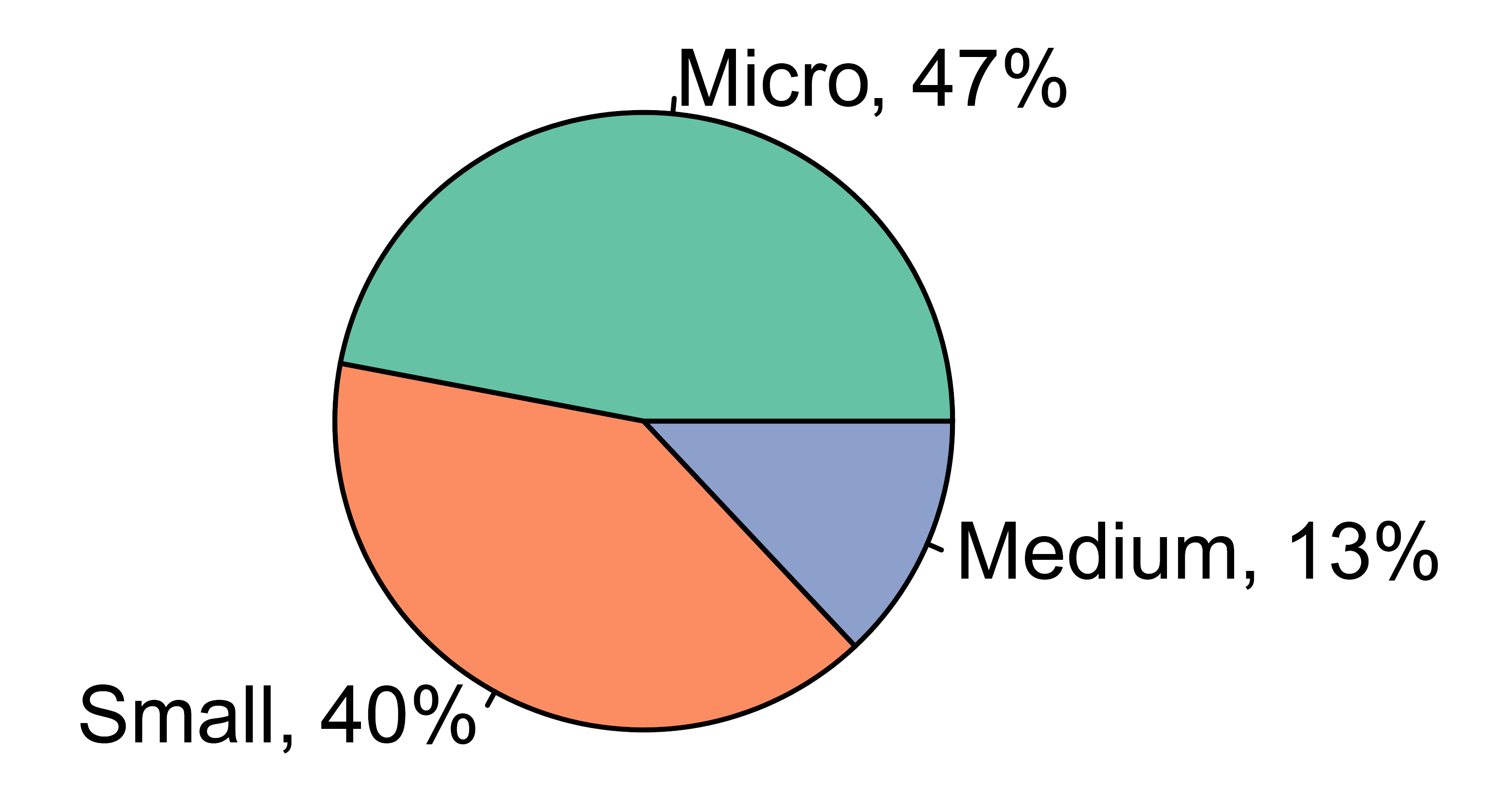}%
\label{FG-size}}
\caption{Demographics of the businesses who took part in the Focus Group Study.}
\label{focus-group-demog}
\end{figure*}

Demographics: A total of 7 females and 8 males took part in this exercise. They represented 9 different sectors, with company sizes ranging from 1 to 62 employees. This data is represented in Figure~\ref{FG-sector} and Figure~\ref{FG-size}. 

\subsection{Recruitment}
Participants were required to be an SME owner or manager. Participants for the think-aloud interviews had previously taken part in stakeholder interviews for the project. They were originally recruited via snowball sampling. The 20 participants for the focus group attended the focus group after either being invited by the workshop host (Kerry local enterprise office or Culture City) or by registering their interest after seeing the flyer advertising the event. The data collected from three participants in the focus groups were omitted, as they were part of larger companies (employees $>250$) and this study focused on small businesses only. Two other participants in the focus group did not complete the risk assessment tool. In total 29 participants completed the risk assessment tool, 14 from the think-aloud interviews and 15 from the focus groups.

The average time to complete the risk assessment tool during the think-aloud activity was 17 minutes and 40 seconds. These participants were given a €50 One4All voucher after participation, as the research team recognise how busy SME owners are and this demonstrated our appreciation to them for taking time out to complete the risk assessment tool. Data collection from these interviews took place from the end of March 2024 to the end of April 2024. 

A time limit of 10 minutes for completing the risk assessment tool was given to the participants in the focus group. The average time for completion was 8 minutes and 34 seconds. These participants did not receive a voucher; however, after the focus group exercise, they received a free workshop on two major areas of risk for SME's: Data Backups and Access Management. The two focus groups took place at the end of April, after all the Think-Aloud interviews had been completed.

Both studies were approved by the University Research Ethics Committee. Approval code: ANON\_HREC-MR-24-007-A.

\subsection{Limitations}
One limitation of our study was the potential inherent bias from recruiting participants by asking them to take part in a cybersecurity study. It is possible the individuals who volunteered to take part already had an interest in cybersecurity. To address this, we included a question asking participants to rate their confidence in the cybersecurity of their business on a scale of 1 to 5. Out of the 19 participants asked, only one rated their confidence at the highest mark of 5, with an average score of 3.12. This suggests that although the participants may be interested in cybersecurity, they are not confident in their knowledge. 

Another limitation in the study is the use of the think-aloud method. Some studies suggest that the presence of a researcher can hinder the outcome of the activity, as some participants may feel self-conscious or awkward~\cite{may2018youtube}. However, the extensive feedback and suggestions received from the participants in the think-aloud interviews, suggest that the participants felt comfortable with the researcher and their presence did not adversely affect the outcome. 

\section{Results}
\subsection{International Risk Assessment Tools}
The goal was to analyse an example nation-state-backed cyber risk assessment tool from each inhabited continent. The reason for nation-state-backed risk assessment tools is to ground our risk assessment tool in real-world best practice examples that countries have currently chosen to deploy. This approach is distinct from the range of theoretical cyber risk assessment tools researchers have developed. We aimed to include a risk assessment tool from at least one country in each inhabited continent. However, finding a nation-state-backed risk assessment tool for Africa and South America proved difficult. Nevertheless, some information regarding risk assessment tools and SME cybersecurity was found for both continents. 

Several African countries provided valuable information and tips regarding cybersecurity. For example, Kenya \cite{Kenya} offers comprehensive cybersecurity guidelines.  Nigeria \cite{Nigeria} also shared important information on cybersecurity, and the government has published a document on how to conduct a cyber risk-assessment, providing links to several risk assessment tools. However, the published material is directed towards financial institutions. In South America, Brazil ~\cite{Brazil} emphasizes the importance of cybersecurity education. The Global Cyber Alliance released their toolkit in Brazilian Portuguese, indication an appetite for cyber knowledge in the region. The toolkit covers device safety, password management, phishing and malware prevention, backup and recovery, and email and reputation protection, with accompanying videos and information. One common theme among all these countries is the recognition of the urgent need to promote and encourage better cybersecurity practices. 

The countries that did have a nationally supported or endorsed risk assessment tool that were included in the analysis were: Belgium~\cite{belgianRAT}, Singapore \cite{SingaporeRAT}, Australia \cite{AustralianRAT} Finland \cite{FinlandRAT}, UK \cite{UKRAT}, and US \cite{USRAT}. 

\subsection{Analysis of existing Risk Assessment Tools}
Once the risk assessment tools were identified, a single researcher completed each risk assessment tool under the following three personas. These personas were created based on 70 stakeholder interviews conducted during the ideation phase of this project: 
\begin{itemize}
 \item Persona 1:  Florist, 5 employees, Has a website set up by a third party, they can just log in/out of it, Do not outsource cybersecurity, Do not have a BCP, Has customer personal data, Deals with third parties (deliveries).
 \item Persona 2: Property/residential sales, 12 employees, Does not train staff on cybersecurity, Outsources IT, No business continuity plan (BCP), Relies a lot on outsourced IT team.
 \item Persona 3: Outsource/service based, 55 employees, Trains staff on cybersecurity, Has a BCP though it is not tested regularly, Has a dedicated IT team/person.
\end{itemize}

The following quantitative details were recorded: Average time taken, number of questions, percentage of questions which included cybersecurity jargon. A score for Ease of Answering the questions was also assigned which was an indicator of who would be capable of answering the questions, where 1 indicates that a regular employee could answer the questions, 2 that the small business owner could answer them, 3 for IT personnel and 4 for a cyber security expert. These details are recorded in Table~\ref{table_existingRATs}. Each risk assessment tool was also compared in terms of: topics covered, availability, ease of access and question format. This is discussed in the following sections.

\begin{table*}
\caption{Comparison of international risk assessment tools}
\label{table_existingRATs}
\centering
\begin{tabular}{|c||c|c|c|c|}
\hline
\textbf{Country} & \textbf{Average time taken} & \textbf{Number of questions} & \textbf{Percentage Jargon} & \textbf{Ease of answering}\\\hline
Belgium & 10 minutes & 8 &0\%&(2) SME Owner\\\hline
Singapore & 2 minutes & 7 & 0\% & (1) Regular employee\\\hline
Australia & 12 minutes & 37 & 3\% & (1) Regular employee\\\hline
Finland & 120 minutes &  338 & 45\% & (4) Cybersecurity professional\\\hline
UK & 9 minutes & 44 & 0\% & (1) Regular employee\\\hline
USA & 10 minutes & 30 & 0\% & (2) SME Owner\\
\hline
\end{tabular}
\end{table*}

\subsubsection{Themes covered}
Open coding using an inductive coding protocol~\cite{gibbs2007thematic,creswell2017research} was conducted to identify the topics covered by each risk assessment tool. The following themes were identified: Passwords, Software Updates, Antivirus/Antimalware, Network Security, Backup Procedure, Administration Rights, Physical Security, Incident Preparedness, Cybersecurity training/Awareness, Workforce Dedication to IT/Cybersecurity, Access Management, and Understanding Data. After identifying the set of themes, the researchers went through each risk assessment tool to identify which country's tools covered which topics. The top two topics were Software Updates and Backup Procedure, which were mentioned by all countries except Singapore. Australia's risk assessment tool covered 9 out of the 12 topics, and Belgium covered 8. Singapore covered the least number of topics. The topics covered by each country's risk assessment tool are indicated in Table~\ref{table_existingRATs_topics}.

\begin{table*}
\setlength{\extrarowheight}{1.5pt}
\caption{Topics covered by Existing Risk Assessment Tools}
\label{table_existingRATs_topics}
\centering
\begin{tabular}{|l||c|c|c|c|c|c|c|c|c|}
\hline
\multicolumn{1}{|c||}{Country} &\multicolumn{1}{p{1.4cm}|}{\textbf{Passwords}} & \multicolumn{1}{p{1.3cm}|}{\textbf{Software Updates}} & \multicolumn{1}{p{1.7cm}|}{\textbf{Anti-virus/ Anti-malware}} & \multicolumn{1}{p{1.3cm}|}{\textbf{Network security}} & \multicolumn{1}{p{1.4cm}|}{\textbf{Backup Procedure}} & \multicolumn{1}{p{1.6cm}|}{\textbf{Adminstrator rights}} & \multicolumn{1}{p{1.3cm}|}{\textbf{Physical Security}} & \multicolumn{1}{p{1.7cm}|}{\textbf{Incident\hphantom{hel} preparedness}} & \multicolumn{1}{p{1.7cm}|}{\textbf{Cybersecurity training/ awareness}}\\\hline
Belgium & \cellcolor{green}{\checkmark}& \cellcolor{green}{\checkmark}& \cellcolor{green}{\checkmark}& \cellcolor{green}{\checkmark}& \cellcolor{green}{\checkmark}& \cellcolor{green}{\checkmark}& \cellcolor{green}{\checkmark}& \cellcolor{green}{\checkmark}&\\\hline
Singapore & &&&&&&& \cellcolor{green}{\checkmark}&\\\hline
Australia & \cellcolor{green}{\checkmark}& \cellcolor{green}{\checkmark}& & \cellcolor{green}{\checkmark}& \cellcolor{green}{\checkmark}& & & \cellcolor{green}{\checkmark}&\cellcolor{green}{\checkmark}\\\hline
Finland & & \cellcolor{green}{\checkmark}& & & \cellcolor{green}{\checkmark}& & & \cellcolor{green}{\checkmark}&\cellcolor{green}{\checkmark}\\\hline
UK & \cellcolor{green}{\checkmark} & \cellcolor{green}{\checkmark} & \cellcolor{green}{\checkmark} & & \cellcolor{green}{\checkmark}&&&&\\\hline
USA &  &  & & & \cellcolor{green}{\checkmark} && \cellcolor{green}{\checkmark} &\cellcolor{green}{\checkmark}& \cellcolor{green}{\checkmark}\\\hline
\end{tabular}
\end{table*}

\subsubsection{Availability and Ease of Access}
Some of the risk assessment tools were easier to locate online than others. The tools from Belgium, Singapore, and Australia were easily sourced via a simple Google search. These tools were also easily accessible and readily available for users.  There was no sign up required for Belgium and Singapore's risk assessment tool, however an Australian Business Number (ABN) was required to begin Australia's risk assessment tool. In contrast, the tools from Finland and the US were more difficult to locate. Finding Finland's risk assessment tool, required an extensive Google search, and when the tool was found, it needed to be downloaded. However, the difficulty in finding the risk assessment tool could be due to the language barrier and limitations in the use of Google Translate to find appropriate search terms. The US risk assessment tool was also difficult to find, and restricts users to completing the assessment only once per device. 

\subsubsection{Question Format}
\begin{figure*}[!t]
\centering
\subfloat[The US Risk assessment tool used a 1-5 scale format as shown above for most questions.]{\includegraphics[height=2in]{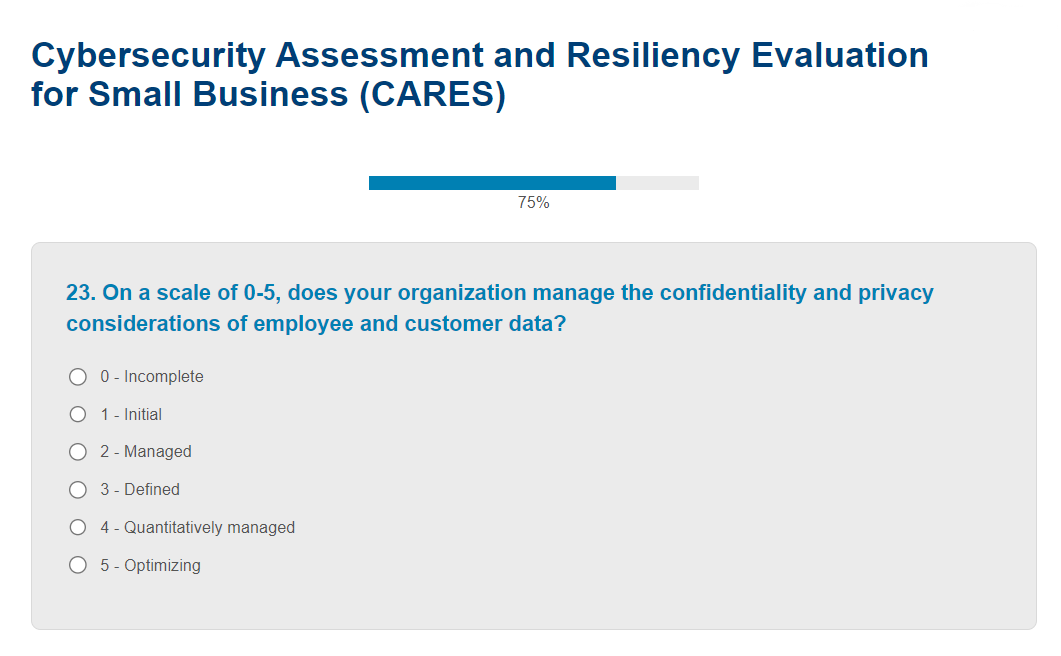}%
\label{fig:exampleUSQ}}
\hfil
\subfloat[The Australian Risk assessment tool used various statement as the answer options for each question.]{\includegraphics[height=2in]{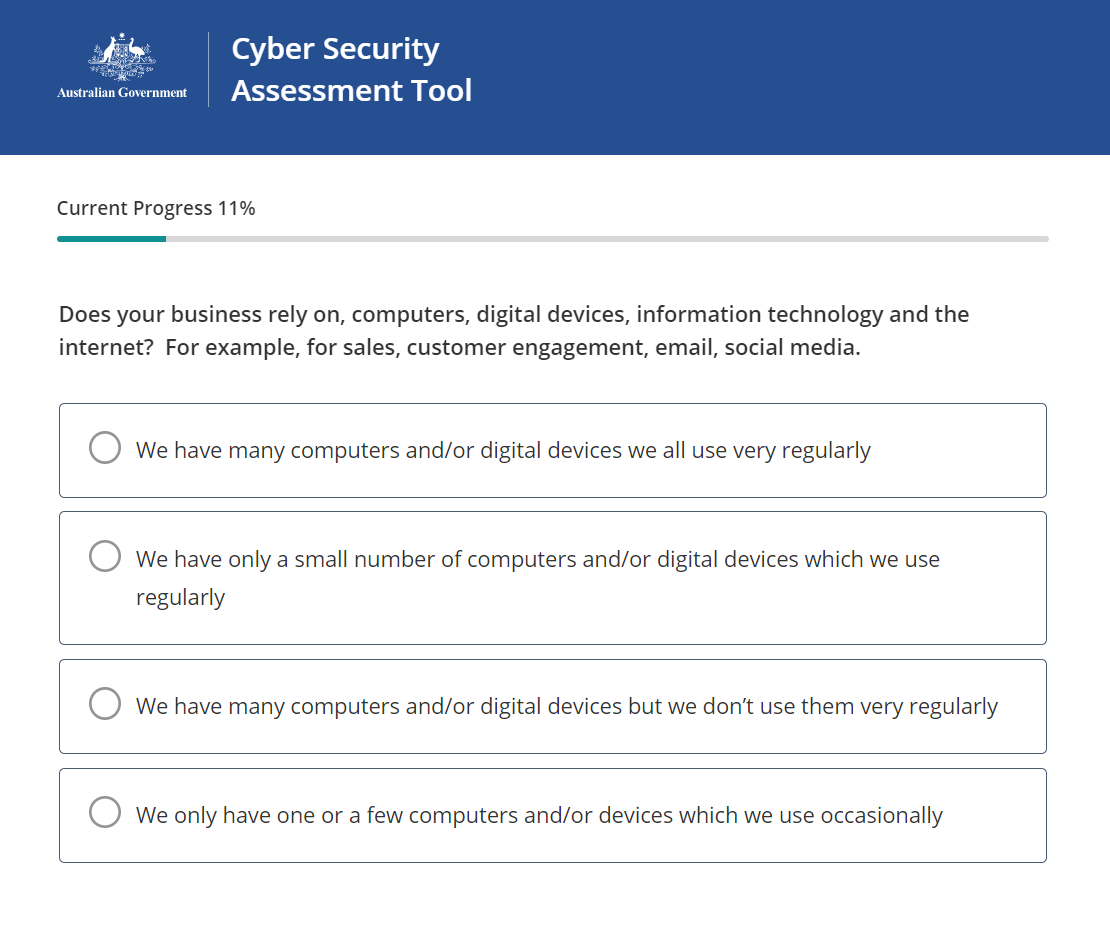}%
\label{fig:exampleOzQ}}
\caption{Example questions from the US and Australian Risk assessment Tools.}
\label{example-qs}
\end{figure*}

For Finland and the US's risk assessment tools, the majority of the questions gave the user a statement and asked them to rank their business on a scale of 1-5 (See Figure~\ref{fig:exampleUSQ}). This approach posed a few issues: the researchers found their attention declining after repeated self-ranking questions, it is also unclear how a user could accurately gauge their position on the scale, leading to guesswork. In comparison, Australia's answer options included 4-6 statements, some with multiple options, allowing users to clearly identify which category they fall into (See Figure~\ref{fig:exampleOzQ}). 

\subsubsection{Language used}
The language used in majority of the risk assessments was easy to understand and accessible to someone with little to no cybersecurity knowledge. The researchers found particularly commendable the option in the UK's survey to say `I don't know what this is' and have a description of the term come up. Australia also took usability into consideration as after most of their questions they give an example or an explanation. These descriptions help users feel more considered and reduce the likelihood of them abandoning the tool due to incomprehensible questions. 

Finland's risk assessment tool used technical language that is better suited to a committed small business willing to take the time to get a highly personalised evaluation of risks. For a risk assessment tool that will be targeted at SMEs, especially micro enterprises, with little to no cybersecurity knowledge, the explanations and non-technical language are a necessity to ensure user accessibility.

\subsubsection{Number of questions}
Originally, we believed that a long, question-heavy risk assessment tool would be non-user-friendly. However, Australia's risk assessment tool at 37 questions, and the UKs at 44 questions, did not feel too lengthy when undertaking the assessment. This could be due to the accessible language used by the tool as both could be completed by a non-cybersecurity expert and were completed by the researcher in under 12 minutes. Finland's Risk Assessment Tool at 338 questions seemed far too time-consuming. Another limitation of Finland's risk assessment tool is that it was not as easy to fill out as the others, due to needing to be downloaded and the use of technical language in most questions. It took the researcher 2 hours in two sittings to complete this risk assessment tool for 1 persona. Combined with the complex cybersecurity language used, it appears this risk assessment tool is more suited to businesses that have some cybersecurity knowledge or expertise. This is not the case for the majority of small businesses~\cite{erdogan2023cybersecurity}. Singapore and Belgium had 7 and 8 questions, respectively. Singapore's questions were focused on the user's time and resources available for cybersecurity training, so the business' broader cyber security risks were not analysed. In Belgium's risk assessment tool, the questions included a variety of topics, so although it was short, a range of cyber risks were assessed. It seems reasonable for a risk assessment tool to contain anywhere from 8-50 questions; if on the lower end, then it must include a variety of topics, and if on the higher end, it should include simple explanations, accessible language and constitute easy to answer questions. 

\subsubsection{Conclusion: what makes an effective small business cyber risk assessment tool}
Considering the factors of each of the risk assessment tools, our initial impressions for a user-friendly risk assessment tool are as follows: easy to locate on the internet after a quick search, readily available for users without the need to create an account or to download it, about 20-50 questions that are well explained, having statements as the answer options and only using rankings of 1-5 minimally, with easy and accessible language used throughout.


\section{Piloting the Risk Assessment Tool}
Based on the analysis of existing risk assessment tools, prior stakeholder engagement with SMEs and best practice literature, the research team developed a first draft of the risk assessment tool. 
 
Two versions of this initial risk assessment tool were created using Microsoft Forms: one for companies with only one employee and another for companies with more than one employee. The questions were similar, but worded differently, with some questions regarding staff being omitted from the `One-Employee' version. 

As described in the Methodology Section~\ref{sec:meth}, the draft risk assessments were piloted with 14 SME business owners or managers in a think-aloud study. Each participant was asked to complete the risk assessment while voicing their thoughts and suggestions throughout. In the below sections we detail the improvements made to the risk assessment tools based on this feedback. Note that simple changes were made after every 3 interviews so that the survey was iteratively improved. For larger suggestions, the study team collated all the recommendations before making a decision regarding the changes that would be implemented.

\subsection{Suggestions implemented}
\subsubsection{Terminology}
There were two risk assessment tools; one for single-employee companies (which was originally stated as `sole-trader') and one for companies with more than one employee. The first change made, after feedback from several participants, was changing the wording of `sole-trader' to `only employee'. This was due to some companies having just one employee but not operating as a sole-trader.

A couple of participants asked for explanations on certain terms. These included: `testing a backup,' `immutable,' `air-gapped,' and `Password Manager'. Explanations were added next to these terms, which proved to be a positive change as the overall average rating from the Pilot interviews on language used was 4.43 out of 5, with one participant saying: `Some unusual words were used but they were then explained, so it was fine'.

\subsubsection{Backups}
There were suggestions made in relation to the data back-up questions, or lack thereof. One participant (P4\_I) suggested the addition of the question: `do you back up your website?'. This feedback was taken on board, and another participant (P6\_I) then suggested changing the wording to `is your website backed up?' as in some cases the website is backed up by another employee or a third party. It was also suggested by P4\_I to add a follow-on question to `do you wipe old devices?' and ask, `is the data on the device backed up before wiping?'.

It was recommended by participants P4\_I, P6\_I, P9\_I, and P13\_I to add in an extra answer option for one of the questions relating to how the user performs data backups. Originally the options were that the data backups were done manually, automatic, or not at all. A combination answer was added, as some back-ups were being completed manually and others automatic. It was also highlighted on the question regarding backup frequency, that different data sets would be backed up at different time frames. To address this, multiple answer options were allowed.

\subsection{Suggestions considered but not implemented}

\subsubsection{Asset tagging}
One participant from a larger SME suggested adding in a question asking whether assets are tagged. Although this is important, the decision was made to not include this question as it was deemed to not be of enough importance for a small business when an inventory of devices will often suffice.

\subsubsection{False questions}
It was prompted to add in a false question regarding passwords asking `how often are passwords changed?'. This advice was not taken, as it was noted that the users were learning as they were completing the risk assessment tool, and this question would lead them to believe that forcing users to change their password regularly is good practice, when it is not~\cite{nist2017}.

\subsubsection{Other}
One question asked `Where do you store this data?' P4\_I suggested that a reference is made to the question number which previously asked what data does the business collect. This was not possible as the question was numbered differently for each user due to branching questions. Instead, the question was rephrased to `where do you store the data that you collect?'.

With regards to encryption, one participant proposed to ask if the user specifically encrypts data, as most software automatically encrypt data. This was deemed to be slightly too advanced for someone with no cybersecurity knowledge to know, so the question remained as `Do you encrypt the data that you store? (for example, do you keep the data in a file that is password protected?)'.

Another thing noted in relation to the password questions, was that a few participants said they did share some passwords but went on to say they had no choice as some software does not allow multiple accounts. For the purpose of research, a follow-on question was added (which will not be included in the final version) asking `What is the reason for sharing passwords?' This question led to valuable advice being given at the workshops, where the participants were taught how to add multiple users to a shared email account, allowing for everyone to have individual log in details. This piece of advice was well received.

\subsection{Written feedback on the Risk Assessment Tool}
After completing the risk assessment tool, each participant was automatically redirected to a feedback survey. 

This simple feedback form included 4 feedback prompts each requesting a rating of 1-5 (5 being the highest) with the option to comment further on the ratings, and a final chance for the participant to provide any other feedback. The rating prompts included: `The language used throughout was understandable (no jargon used),' `The tool was easy to use,' `The questions were relevant to my business,' and `It was a beneficial exercise.' The average ratings for each statement, respectively, were: 4.43, 4.86, 4.43, 4.71. While these are likely higher than they would be if the participant had used the tool without a researcher present, the comments showed a strong support for the value of the risk assessment tool. 

All participants found the language used in the tool easy to understand, with an average score of 4.33 out of 5. One of the first participants commented ``Some jargon I was not sure what it meant ...immutable air gapped and password manager.'' This issue was rectified by the research team, which is reflected in the comment of one of the last participants who said, ``I like the fact there was clarification for some of the unknown terminology to me.''

All participants found the tool easy to use, with an average score of 4.86 out of 5. Some comments include: ``Simple survey style, very easy to use... and quick!'' ``Very easy to use and navigate'' ``easy to use and well organized.'' Two comments suggested the addition of more options: ``Add some supplementary questions when applicable or not applicable,'' ``Easy to use, but more options needed. We are partially between many options.'' This feedback was taken on board and more answer options were added to be more inclusive.

Most of the participants found that the questions were relevant to their business, with an average score of 4.43 out of 5. Some participants found it was completely relevant to them: ``Hugely relevant to any business,'' ``100\% relevant.'' While others highlighted how the questions were relevant but also eye opening: ``Very relevant to my business and gave me lots of food for thought!'' ``Make me question additional items and as such was very relevant''. Some participants found that not all questions were relevant to their business: ``I have a small company so some not as relevant as only 2 employees,'' ``I operate a taxi so most questions aren't relevant''.

The average rating for the exercise being beneficial was 4.71 out of 5. Despite some participants commenting that they felt the questions were not fully relevant to them, they still found the exercise beneficial, with both participants scoring this at a 5. One of these participants commented: ``It has made me aware that I need Cyber security training and a plan in place in case of an attack.'' This was the main consensus for this section, with most of the participants saying the exercise was beneficial and made them aware of their cybersecurity risk.


\section{Focus group study}
Once the final changes were made to the risk assessment tool based on the feedback from the think-aloud pilot study, the new risk assessment tool was created using Microsoft Forms. Again, two versions were created, one to cater for companies with only one employee and one for companies with more than one employee. 

A new set of participants were invited to attend a focus group and workshop on cyber resilience for small businesses. The workshop began with each participant completing the online risk assessment tool for their own business. This was followed by focus group discussions with the generic prompt ``How did you find that exercise?''. The researchers encouraged open discussion among the participants about their thoughts on using the tool and did not interject with any additional prompts or directions but allowed the conversation to flow among the participants. The points made in the discussion were transcribed by the researchers and analysed afterwards.

\subsection{Focus Group Discussions}
\subsubsection{Thought-provoking}
The main feedback received from the focus groups was how the risk assessment tool was a thought-provoking activity. One participant said: ``stuff there, you look at it and go, oh that's really basic, I should be doing that but I never thought of it''. At the same focus group, another participant said: ``It makes you think, when you are seeing more no's than yeses of what you have and don't have.'' This was a positive piece of feedback to hear as it confirms the hope that users would be learning as they completed the risk assessment tool, hence why the suggestion of adding in the false question regarding changing passwords was dismissed.

\subsubsection{Perceived Challenges and Fears}
Some participants relayed how they feel cybersecurity is too big a task. One participant said, ``I try to collect less data because then I don't have to manage it.'' This feedback suggests that some SMEs are aware of their risks but are using avoidance as a coping mechanism rather than learning how to manage their cybersecurity. Several participants mentioned how they put their trust into their software systems. After completing the risk assessment tool, one participant said they realised how reliant they are on the software systems. Another said they worry about the data they collect and that all their trust is on the software suppliers. The perception of cybersecurity as an overwhelming task suggests that giving a prioritised checklist would help some SMEs as it would feel less intimidating and give them the opportunity to have less reliance on their software systems and more confidence in themselves. 
 
\subsubsection{Suggestions}
The discussion with the focus group led to more feedback than suggestions. It is possible that the group discussion environment and the researcher's hierarchy style of workshop format was not conducive to direct critique of the tool. For this reason, for suggestions, we turn mostly to the anonymous feedback all participants submitted directly through the risk assessment tool portal. 

There was one query while participants completed the tool from an individual who wanted to clarify the meaning of the questions which involved assigning a star rating. They also included this comment as part of their written feedback and text has since been added to the star rating questions explaining that 1 star is the lowest and 5 stars is the highest.

\subsection{Written feedback for Risk Assessment Tool Version 2}
 
 The feedback section at the end of the risk assessment tool remained the same as for the pilot version. For the 4 feedback prompts: `The language used throughout was understandable (no jargon used)', `The tool was easy to use', `The questions were relevant to my business', and `It was a beneficial exercise' the ratings were: 4.33, 4.6, 4.47, and 4.6. These are slightly lower than those recorded in the think aloud pilot which likely reflects the anonymity of these responses where before the participants were actively on a call with the researcher while completing the tool. The score for `The questions were relevant to my business' increased minutely from 4.43 to 4.47. There was less written feedback from the focus group participants than that of the think aloud participants. The main recommendations for improvement were ``Air gapped and immutable explanations were good but if there were an asterix used in original text be better'', one person was unclear on the star scoring methodology, i.e. that 5 stars is a high rating and 1 star low. Both of these accessibility considerations were updated in the final version of the risk assessment form.

All of the participants found the language easy to understand, giving an average score of 4.33 out of 5. Participants highlighted the benefits of explanations being given with some terms: ``Very straightforward and explanations given in certain areas'', ``Anything that I was not familiar with was explained in the same question''.

Participants also found the tool easy to use, rating it at 4.6 out of 5. One person gave a 3/5 rating. This participant had trouble connecting at the beginning due to network issues. All other participants commented on how easy the tool was to use: ``Easy to use'', ``Easy to use and in a language that was approachable''.

All participants gave either a 4 or 5 rating for the relevance of the questions to their business and for the value of completing the risk assessment tool. The average score for the relevance of the questions was 4.47 out of 5, which was higher than that of the think-aloud study. As with the think-aloud pilot group, some participants highlighted how the questions were eye-opening for them: ``Made me think of more things I should be doing''. One participant felt the questions were not relevant at the moment but recognised that they will be relevant in the future: ``As I'm only starting my business some of the questions might not apply to me at the moment but will in the future''. The average score for the benefit of completing the exercise was 4.6/5. Many participants used this comment section to express how they have realised what they need to do: ``I need more awareness'', ``I need to up my game''. Raising awareness of cyber risks to small business owners in an inclusive and empowering manner is the core goal of a small business risk assessment. Participants commented: ``Definitely gave me a lot to think about'' and ``Good to see what is being asked to know what needs to be looked at in our business''.

\section{Key Areas of Risk for SMEs}
This final section of the paper highlights the core cyber risks of Irish small businesses informed by the anonymised results of the risk assessment responses. \textbf{Areas of high risk among Irish SMEs include lack of knowledge and compliance with EU GDPR regulations, limited data protection and backup procedures, ad-hoc or absent cybersecurity training for employees or owners, and no cyber-incident response plan or business continuity plan.} For each risk area, we indicate below the NIST Cybersecurity Framework (CSF) function it relates to.

\subsection{GDPR Obligations}
A concerning gap in GDPR compliance has been identified among participants, with 26 out of 29 reporting that they collect personal data. However, 8 participants (28\%) admitted, ``I did not know this until now,'' when asked if they were aware of their obligation under GDPR to report a personal data breach to the Data Protection Commissioner within 72 hours of becoming aware of the breach. This lack of awareness is a significant issue, as non-compliance with GDPR reporting requirements can result in severe sanctions for business owners. According to the NIST Cybersecurity Framework (CSF) (National Institute of Standards and Technology, 2024) understanding and adhering to regulatory requirements is crucial for maintaining a robust cybersecurity posture (GV.OC-03).

\subsection{Data Risk}
Data protection practices among participants show significant vulnerabilities. When asked about encryption, 16 out of 29 participants (55\%) responded ‘No’ or ‘I’m not sure’ if they encrypt the data they store. Additionally, 11 participants (38\%) lacked a formal backup procedure, and among those who did, 8 were uncertain about the frequency of their backups. Only 2 participants reported having immutable or air-gapped backups. Furthermore, 67\% of participants indicated they do not regularly test their backups. According to the NIST CSF, effective data protection (PR.DS) and backup strategies (PR.DS-11) are essential for mitigating risks and ensuring data resilience. 

\subsection{Cybersecurity Training and Awareness }
The survey revealed a substantial deficiency in cybersecurity training and awareness among participants. Sixteen out of 29 participants (55\%) reported that neither they nor their staff have ever engaged in cybersecurity training. Only one participant reported weekly engagement in cybersecurity training, while 3 participants indicated annual training. The remainder stated that training occurs on an ad hoc basis or only upon hire. The NIST CSF emphasizes that regular cybersecurity training and awareness are critical for fostering a culture of security and resilience within an organization (PR.AT). 

\subsection{Incident Preparedness}
Incident preparedness is another critical area where participants fell short. Only 8 out of 21 participants (38\%) required their staff to report suspicious activities or security incidents. Alarmingly, 20 participants (69\%) stated they would not know how to respond if a cyber incident occurred. Of the participants who had a cyber incident response plan, half had never tested it. Additionally, 21 out of 29 participants (72\%) lacked a Business Continuity Plan, and only 2 regularly tested and reviewed their plans. The NIST CSF highlights the importance of incident response planning and regular testing to ensure organizational readiness and resilience (RS).

\subsection{Other Areas of Risk }
Several additional risk areas were identified among participants. Out of 12 participants managing their website’s cybersecurity, 5 (42\%) lacked confidence in their ability to do so. Furthermore, 21 out of 27 participants (78\%) reported using personal devices for work, with no clear policies in place. This statistic underscores the need for comprehensive device management policies, as recommended by the NIST CSF (ID.AM). Additionally, 7 participants (24\%) were unsure if they had cyber insurance, indicating a lack of awareness about their insurance policies. Thirteen out of 21 participants (62\%) admitted to sharing passwords, highlighting poor password management practices.

\section{Discussion}

The development of our risk assessment tool has unveiled several key insights which has allowed us to enhance its efficacy and relevance for SMEs on an iterative basis. 

It was surprising that there was no clear, common structure in the international risk assessment tools. Although their intentions seemed the same, the risk assessment tools themselves were vastly different in their look and feel. Despite the wide variety of literature on how to create a risk assessment tool, this literature did not seem to be applied in the design of the analysed tools. 

A second insight from our iterative design process was that by virtue of just filling in the risk assessment tool, participants were educated on their cybersecurity risk gaps. This emphasizes the value of a risk assessment tool and the role it plays in understanding cybersecurity gaps. A big part of cybersecurity risk is understanding your risk areas. If people are unable to appreciate their level of risk, they will be unable to improve. 

Small businesses have a variety of cybersecurity risks, often unbeknownst to them~\cite{erdogan2023cybersecurity}. In the pilot studies and the focus groups our participants expressed an appreciation for the clear identification of clear risk concepts. SME risk areas ranged from GDPR non-compliance, data storage and backup practices to lack of cybersecurity awareness, policies or training. 

 It is difficult to quantify the value of a cybersecurity risk assessment tool. Our success metrics were based on qualitative feedback and quantitative feedback via a survey at the end of the risk assessment tool. We asked participants to give a rating of 1-5 on the following statements: ‘The language used throughout was understandable (no jargon used)’, ‘The tool was easy to use’, ‘The questions were relevant to my business’, and ‘It was a beneficial exercise’. These are indicators for what we saw as success criteria for our risk assessment tool. However, the success of the tool can be more clearly seen through its demonstrated ability to empower rather than intimidate small business owners on the topic of cybersecurity ``I need to up my game''.

 Our goal was to create an accessible, user-friendly risk assessment tool. This was made possible due to the user-centred approach that was taken. Incorporating SMEs in the development of the tool allowed us to enhance the tool based on their feedback. The think-aloud interviews were a highlight in the process of creating the risk assessment tool. Both from direct suggestions from the participants and from interview insights the researchers gathered. Through their feedback and our observation, we were able to enhance the risk assessment tool further by using more user-friendly language and adding more options to ensure inclusivity was being achieved.

The final risk assessment tool is available on our website~\cite{ourRAT}.

\section{Conclusion}
This study addresses the critical need for accessible cybersecurity tools tailored to SMEs. SMEs have limited time and budget to dedicate to cybersecurity. Despite this, they are still aware of the need to have good cybersecurity practices but are unsure how to go about this task. 

Through our research, we developed a cyber risk assessment tool specifically designed for SME owners with little to no cybersecurity knowledge. Although the investment from SMEs is minimal, requiring only 10 minutes to use the tool, the payoff is substantial as it has demonstrated that it improved their knowledge of their cyber risks. 

In conclusion, this study fills a vital gap in the current cybersecurity landscape by providing an Irish SME-specific risk assessment tool. This study has also set a foundation for future research and development in this area. By supporting Irish SMEs and providing them with accessible and user-friendly tools, we can significantly strengthen their defenses against cyber threats and increase their cyber resilience.

\subsubsection*{Future work} As requested by several participants in our study, the next step is to generate a cyber risk action plan tailored to the individual needs of an SME based on their answers in the risk assessment tool. This will be the next step in the research study.

\ifCLASSOPTIONcompsoc
 \section*{Acknowledgments}
\else
 \section*{Acknowledgment}
\fi

The authors would like to thank Anon.



%

\bibliographystyle{IEEEtran}
\bibliography{bib}

\end{document}